\documentstyle[preprint,aps]{revtex}
\tightenlines

\begin{document}
\draft
\title{SELF - AFFINITY OF ORDINARY LEVY MOTION,\\
SPURIOUS MULTI - AFFINITY \\
AND \\
PSEUDO - GAUSSIAN RELATIONS}
\author{A. V. Chechkin and V. Yu. Gonchar}
\address{Institute for Theoretical Physics\\
National Science Center ``Kharkov Institute of Physics and Technology`` \\
Akademicheskaya St.1, Kharkov 310108, Ukraine }
\address{and Institute for Single Crystals\\
National Academy of Sciences of Ukraine \\
Lenin ave. 60, Kharkov 310001, Ukraine}
\date{\today }
\maketitle

\begin{abstract}
\begin{center}
The ordinary Levy motion is a random process whose stationary independent
increments are statistically self - affine and distributed with a stable
probability law characterized by the Levy index $\alpha ,$ 0$<\alpha <$2.
The divergence of statistical moments of the order $q>\alpha $ leads to an
important role of the finite sample effects. The objective of this paper is
to study the influence of these effects on the self - affine properties of
the ordinary Levy motion, namely, on the ''1/$\alpha $ laws'', that is, time
dependence of the $q$-th order structure function and of the range.
Analytical estimates and simulations of the finite sample effects clearly
demonstrates three phenomena: spurious multi - affinity of the Levy motion,
strong dependence of the structure function on the sample size at $q>\alpha $%
, and pseudo - Gaussian behavior of the second - order structure function
and of the normalized range. We discuss these phenomena in detail and
propose the modified Hurst method for empirical rescaled range analysis.
\end{center}
\end{abstract}

\pacs{PACS number(s): 02.50.-r, 05.40.+j}


\section{Introduction.}

By Levy motions, one designates a class of random functions, which are a
natural generalization of the Brownian motion, and whose increments are
stationary, statistically self-affine and stably distributed in the sense of
P. Levy [1]. Two important subclasses are (i) $\alpha $- stable processes,
or the ordinary Levy motion (oLm), which generalizes the ordinary Brownian
motion, or the Wiener process, and whose increments are independent, and
(ii) the fractional Levy motion, which generalizes the fractional Brownian
motion and has an infinite span of interdependence.

The theory of processes with independent increments was developed beginning
from the Bachelier's paper [2] concerning Brownian motion. However, the
rigorous construction of this process and studies of properties of its
trajectories were undertaken by Wiener [3]. The modern presentation of the
general theory of processes with independent increments is contained in [4].
The theory of the processes with independent increments possessing stable
distributions has begun its history from the already cited work [1] and,
later on, was developed by other prominent mathematicians. In particular,
the properties of extremes of $\alpha $- stable symmetric processes were
studied in [5]. The geometric properties of their trajectories were
considered in [6]. The monograph [7] contains a modern presentation of the
theory of $\alpha $- stable processes.

The Levy random processes play an important role in different areas of
application for at least two reasons.

The first one is that the Levy motion can be considered as a generalization
of the Brownian motion. Indeed, the mathematical foundation of the
generalization are remarkable properties of stable probability laws. From
the limit theorems point of view, the stable distributions are a
generalization of widely used Gaussian distribution. Namely, stable
distributions are the limit ones for the distributions of (properly
normalized) sums of independent identically distributed (i.i.d.) random
variables [8]. Therefore, these distributions (like the Gaussian one) occur,
when the evolution of a physical system or the result of an experiment are
determined by the sum of a large number of identical independent random
factors. An important distinction of stable probability densities is the
power law tails decreasing as $\left| x\right| ^{-1-\alpha }$ , $\alpha $ is
the Levy index, 0 $<\alpha $ $<$2. Hence, the distribution moments of the
order $q\geq \alpha $ diverge. In particular, stably distributed variables
possess a non-finite variance.

The second reason for ubiquity of the Levy motions is their remarkable
property of scale - invariance. From this point of view the Levy motions
(like the Brownian ones) belong to the so - called fractal random processes.
Indeed, the objects in nature rarely exhibit exact self - similarity ( like
the Von Koch curve), or self - affinity. On the contrary, these properties
have to be understood in a probabilistic sense [9,10]. The random fractals
are believed to be widely spread in nature. A coastline is a simple example
of statistically self - similar object [9], as well as the spot of the
Chernobyl contamination in the nearest zone [11]. On the contrary, the trace
of the Brownian motion is statistically self - affine. Several numerical
algorithms were developed in order to simulate fractional Brownian motion
[10,12 - 15]. They allow one to model many highly irregular natural objects,
which can be viewed as random fractals [10]. The traces of the Levy motions
are also statistically self - affine, therefore, one may expect that they
are also suited for modeling and studies of natural random fractals.

The stable distributions and the Levy, or Levy - like, random processes are
widely used in different areas, where the phenomena possessing scale
invariance (in a probabilistic sense) are viewed or, at least, can be
suspected, e.g., in economy [16 - 19], biology and physiology [20],
turbulence [21] and chaotic dynamics [22], solid state physics [23], plasma
physics [24], geophysics [25] etc. In this respect, the problems connected
with experimental data processing are of great importance. It is also
necessary to develop different numerical algorithms which allow one to
simulate Levy motion with the given statistical properties. This, in turn,
allows one to improve the methods aimed at analysis and interpretation of
experimental data. Recently, three models of the oLm were proposed [26].
They can be interpreted as ''difference schemes'' to approximate the
evolution equation for the distribution density of the ordinary Levy motion.
In our paper we employ the different approximation, which is based on using
Gnedenko limit theorem along with the inversion method for generating random
variables. With the help of this approximation we study the consequences of
self - affinity of the motion, namely, the time dependence of the structure
functions and of the range. We show that the finiteness of the sample size
plays an essential role when the consequences of self - affinity are
considered. We demonstrate both analytically and numerically, that the
finiteness of the sample size violates self - affinity, thus giving rise to
spurious multi - affinity of the oLm. Further, the second order structure
function and the normalized range (just these quantities are widely used in
processing a huge number of various experimental data [12,27]) possess
spurious, ''pseudo - Gaussian'' time - dependence. This circumstance allows
one to suggest that at estimating the second order structure function and
normalized range from experimental data the ''Levy nature'' of them can be
easily masked. In order to avoid pseudo - Gaussianity when studying the
range, we propose a modified Hurst method for rescaled range analysis.

The paper is organized as follows. In Sec. 2 we discuss the property of self
- affinity and its consequences. In Sec. 3 we propose a model for numerical
simulation of the oLm. In Sec. 4 the effects of a finite sample size are
studied. In Sec. 5 we present numerical results. Finally, the conclusions
are exposed in Sec. 6.

\section{ Self - affinity of the ordinary Levy motion}

Let us proceed with the self - affine properties of the ordinary Levy motion
denoted below as $L_\alpha \left( t\right) $. In this paper we restrict
ourselves by symmetric stable distributions. The characteristic function of
the oLm increments is [4,7]

\begin{equation}
\hat p_{\alpha ,D}(k,\tau )\equiv \left\langle \exp \left[ ik\left( L_\alpha
(t+\tau )-L_\alpha (t)\right) \right] \right\rangle =\exp (-D\left| k\right|
^\alpha \tau )  \eqnum{1}
\end{equation}

Here $\alpha \,$is the Levy index, 0 $<\alpha \leq $ 2, and D $>$ 0 is a
scale parameter. The probability density

\begin{equation}
p_{\alpha ,D} (x,t)=\int\limits_{-\infty }^{\infty }\frac{dk}{2\pi } \exp
(-ikx)\hat{p} _{\alpha ,D} (k,t)  \eqnum{2}
\end{equation}

is expressed in terms of elementary functions in two cases:

(i) ordinary Brownian motion, $L_\alpha (t)\equiv B(t)$ , which has $\alpha
=2$ and the probability density

\[
p_{2,D}\left( x,t\right) =\frac 1{\sqrt{4\pi Dt}}\exp \left[ -\frac{x^2}{4Dt}%
\right] 
\]
,

and

(ii) ordinary Cauchy motion, which has $\alpha =1$ and the probability
density

\[
p_{1,D}\left( x,t\right) =\frac{Dt/\pi }{D^2t^2+x^2} 
\]
.

At $\left| x\right| \rightarrow \infty $ the probability densities of the
oLm have power law tails,

\[
p_{\alpha ,D}(x,t)\propto \frac{Dt}{\left| x\right| ^{1+\alpha }} 
\]
.

The increments of the oLm are stationary in a narrow sense,

\begin{equation}
L_\alpha (t_1+\tau )-L_\alpha (t_2+\tau )\stackrel{d}{=}L_\alpha
(t_1)-L_\alpha (t_2),  \eqnum{3}
\end{equation}

and self-affine with the parameter $H=1/\alpha $, that is, for an arbitrary $%
\kappa >$ 0

\begin{equation}
L_\alpha (t+\tau )-L_\alpha (t)\stackrel{d}{=}\left\{ \kappa ^{-1/\alpha
}\left[ L_\alpha (t+\kappa \tau )-L_\alpha (t)\right] \right\} ,  \eqnum{4}
\end{equation}

where $\stackrel{d}{=}$ implies that the two random functions have the same
distribution functions. The exponent $1/\alpha $, by analogy with the theory
of fractional Brownian motion [28], is named the Hurst index of the ordinary
Levy motion.

We consider two corollaries of Eqs.(1) - (4), which, again by analogy with
the definitions of Ref. [28], may be called ''$1/\alpha $ laws'' for the
structure function and for the range.

We first consider the structure function of the oLm. A ''$1/\alpha $ law''
for the structure function can be stated as follows:

for all 0 $<\alpha $ $<$ 2

\begin{equation}
S_q(\tau ,\alpha )=\left\langle \left| L_\alpha (t+\tau )-L_\alpha
(t)\right| ^q\right\rangle =\left\{ 
\begin{array}{c}
\tau ^{q/\alpha }V(q;\alpha ),\ \quad 0\leq q<\alpha \\ 
\infty ,\ \quad q\geq \alpha
\end{array}
\right. ,  \eqnum{5}
\end{equation}

where

\begin{equation}
V(\mu ;\alpha )=\int\limits_{-\infty }^\infty dx_2\left| x_2\right| ^\mu
\int\limits_{-\infty }^\infty \frac{dx_1}{2\pi }\exp (-ix_1x_2-\left|
x_1\right| ^\alpha ),  \eqnum{6}
\end{equation}

whereas for $\alpha =2$ (ordinary Brownian motion)

\begin{equation}
S_q(\tau ;2)=\tau ^{q/2}V(q;2)  \eqnum{7}
\end{equation}

for an arbitrary $q$.

Equations (5) - (7) have a direct physical consequence for description of an
anomalous diffusion. Indeed, for the ordinary Brownian motion the
characteristic displacement $\Delta x(\tau )$ of a particle may be written
in terms of the second order structure function as

\begin{equation}
\Delta x(\tau )=S_2^{1/2}(\tau ;2)=\sqrt{2}\tau ^{1/2},  \eqnum{8}
\end{equation}

where the prefactor $\sqrt{2} $ is simply $V^{1/2} (2;2)$ . One may note
from Eqs.(6), (7) that, for the normal diffusion

\begin{equation}
S_q^{1/q}(q;2)\propto \tau ^{1/2}  \eqnum{9}
\end{equation}

at any $q$ and, thus any order of the structure function may serve as a
measure of a normal diffusion rate:

\begin{equation}
\Delta x(\tau )\approx S_q^{1/q}(q;2)\propto \tau ^{1/2},  \eqnum{10}
\end{equation}

if one is interested in time - dependence of the characteristic
displacement, but not in the value of the prefactor. We remind that usually
just the time - dependence, but not the prefactor, serves as an indicator of
normal or anomalous diffusion [23]. In analogy with Eqs. (9), (10) it
follows from Eqs. (5), (6) that the quantity $S_q^{1/q}(\tau ;\alpha )$ at 0 
$<\alpha $ $<$ 2 and any $q$ $<\alpha $ can serve as a measure of an
anomalous diffusion rate:

\begin{equation}
\Delta x(\tau )\approx S_q^{1/q}(q;\alpha )\propto \tau ^{1/\alpha },\quad
0<q<\alpha <2.  \eqnum{11}
\end{equation}

Here we have the case of a fast anomalous diffusion, or hyperdiffusion.

The second corollary of Eqs. (1) - (4) is for the range of the oLm. A ''$%
1/\alpha $ law'' for the range can be stated as follows:

\begin{equation}
R(\tau )=\sup_{0\leq t\leq \tau }L_\alpha (t)-\inf_{0\leq t\leq \tau
}L_\alpha (t)\stackrel{d}{=}\tau ^{1/\alpha }R(1).  \eqnum{12}
\end{equation}

For the ordinary Brownian motion $\tau ^{1/2}R\left( \tau \right) $ has a
distribution independent of $\tau \,$[29].

For the oLm, $0<\alpha <2$, the statistical mean of the range,

$\left\langle R(\tau )\right\rangle $ , behaves as $\tau ^{1/\alpha }$ at $%
1<\alpha <2$, and turn to infinity at $0<\alpha \leq 1$.

Both ''$1/\alpha $ law'' are studied in numerical simulation in Sec. 5.
However, at the end of this Section we point once more to an important
property of the moments of stable distributions with the Levy index $\alpha
< $ 2: the moments of the order $q\geq \alpha $ diverge. This property, in
turn, manifests itself in divergence of the $q$ - th order structure
function at $q\geq \alpha $ , see Eq.(5), and of the mean of the range at $%
\alpha \leq 1$. However, at experimental data processing both quantities are
finite due to the finiteness of a sample size. In fact, one may expect that
in the case of the Levy motion the finiteness of a sample size has a
stronger influence on the results than in the case of the Brownian motion.
Therefore, the estimates of such influence are needed. We study this problem
in Sec. 4 before discussing results of numerical simulation.

\section{ A simple way to approximate ordinary Levy motion.}

The process of constructing approximation to the oLm can be divided into two
steps.

Step 1. At the first step we generate random sequence of i.i.d. random
variables possessing stable probability law. These variables play the role
of increments of the oLm having the Levy index , $0<\alpha <2$. The value $%
\alpha =2$ corresponds to the ordinary Brownian motion, hence in this case
the sequence of independent increments is generated with the use of a
standard Gaussian generator. Since in Ref. [12] the sequence generated by
the Gaussian generator is called ''approximate discrete - time white
Gaussian noise'', we call the sequence generated at $0<\alpha <2$
''approximate discrete - time white Levy noise''.

We generate approximate white Levy noise possessing characteristic function

\begin{equation}
\hat p_{\alpha ,D}(k)=<\exp (ikx)>=\exp (-D\left| k\right| ^\alpha ). 
\eqnum{13}
\end{equation}

At $0<\alpha <2$ they have power law asymptotic tails [8],

\begin{equation}
p_{\alpha ,D}(x)\propto D\frac{\Gamma (1+\alpha )\sin (\pi \alpha /2)}{\pi
\left| x\right| ^{1+\alpha }},\quad x\rightarrow \pm \infty .  \eqnum{14}
\end{equation}

In the literature there exist different algorithms for generating random
variables distributed with stable probability law. We only mention two
recently proposed schemes, which use the combinations of random number
generators [30] and the family of chaotic dynamical systems with broad
probability distributions [31], respectively. However, we believe that the
ways of generating stably distributed variables and, then the Levy motion
are not exhausted, and various simulation models are needed, each of them
may appear to be useful when studying some particular problem. In this paper
we propose a simple approximation based on the Gnedenko limit theorem along
with the method of inversion. Indeed, among the methods of generating random
sequence with the given probability law $F(x)$ the method of inversion seems
most simple and effective [32]. However, it is well-known fact that its
effectiveness is limited by the laws possessing analytic expressions for $%
F^{-1}$, hence the direct application of the method of inversion to the
stable law is not expedient. In this connection, it is natural to exploit an
important property of stable distributions. Namely, such distributions are
limiting for those of properly normalized sums of i.i.d. random variables
[8]. To be more concrete, we generate the needed random sequence in two
steps. At the first one we generate an ''auxiliary'' sequence of i.i.d.
random variables $\left\{ \xi _j\right\} $ , whose distribution density $%
F^{\prime }(x)$ possesses asymptotics having the same power law dependence
as the stable density with the Levy index has, see Eq.(14). However,
contrary to the stable law, the function $F(x)$ is chosen as simple as
possible in order to get analytic form of $F^{-1}$. For example,

\begin{equation}
F(x)=\frac 1{2(1+\left| x\right| ^\alpha )},\quad x<0,  \eqnum{15}
\end{equation}

\[
F(x)=1-\frac 1{2(1+x^\alpha )},\quad x>0. 
\]

Then, the normalized sum is estimated,

\begin{equation}
X=\frac 1{am^{1/\alpha }}\sum\limits_{j=1}^m\xi _j,  \eqnum{16}
\end{equation}

where

\[
a=\left( \frac \pi {2\Gamma (\alpha )\sin (\pi \alpha /2)}\right) ^{1/\alpha
}. 
\]

According to the Gnedenko theorem on the normal attraction basin of the
stable law [8], the distribution of the sum (16) is then converges to the
stable law with the characteristic function (13) and $D=1$. It is reasonable
to generate random variables having stable distribution with the unit $D$,
with a consequent rescaling, if necessary. Repeating $N$ times the above
procedure, we get a sequence of i.i.d. random variables $\left\{
X(t)\right\} ,\quad t=1,2,...,N$ . This is an approximate discrete - time
white Levy noise.

In the top of Fig.1 the probability densities $p(x)$ for the members of the
sequence $\left\{ X(t)\right\} $ ($m=30$ in Eq.(16)) are depicted by black
points for (a) $\alpha =1.0$, and (b) $\alpha =1.5$. The functions $%
p_{\alpha ,1}(x)$ obtained with the inverse Fourier transform, see Eq.(13),
are shown by solid lines. In the bottom of Fig.1 the black points depict
asymptotics of the same probability densities in log-log scale. The solid
lines show the asymptotics given by Eq.(14). The examples presented
demonstrate a good agreement between the probability densities for the
sequences $\left\{ X(t)\right\} $ obtained with the use of the numerical
algorithm proposed and the densities of the stable laws.

We would like to stress that a certain merit of the proposed model is its
simplicity. It is entirely based on classical formulation of one of the
limit theorems and can be easily generalized for the case of asymmetric
stable distributions. It is also allows one, after some modifications, to
speed up the convergence to the stable law. These problems, as well as the
comparison with the other algorithms seems to be the subject of a separate
paper.

Step 2. With using approximate discrete - time white Levy noise $X(t)$ the
approximation to the oLm is defined by

\begin{equation}
L_\alpha (t)=\sum\limits_{\tau =1}^tX(\tau )  \eqnum{17}
\end{equation}

In Fig.2 the approximate white Levy noises obtained with the numerical
algorithm proposed are depicted by thin lines at 4 different Levy indexes.
The thick lines depict the sample paths, or the trajectories, of the
approximation to the oLm. It is clearly seen that with the Levy index
decreasing, the amplitude of the noise increases. The large outliers lead to
large ``jumps'' (often named as ''Levy flights'') on the trajectory.

\section{ Effects of limited sample size.}

In this Section we discuss what is the time dependence of the $q$ - th order
structure function and the range, which are estimated from experimental data.

We first give an estimate for the mode of the maximum value.

Suppose we have a sequence $\{X(t)\}$, t = 1,2,...,N, of i.i.d. random
variables possessing stable probability density $p_{\alpha ,1}\left(
x\right) $ and cumulative probability $P_{\alpha ,1}(X\leq
x)=\int\limits_{-\infty }^xdu\;p_{\alpha ,1}(u)$ . Then, $P_{\alpha ,1}^N$,
is the probability that all $N$ terms of the sequence are less than $x$.
This, in turn, implies that $P_{\alpha ,1}^N$ is the probability that the
maximum value of $N$ terms is less than $x$. Therefore,

\begin{equation}
\varphi _N(x)=\left( P_{\alpha ,1}^N(x)\right) ^{\prime }=NP_{\alpha
,1}^{N-1}(x)p_{\alpha ,1}(x)  \eqnum{18}
\end{equation}

is the probability density of the maximum value in the sample consisting of $%
N$ terms. Let $X_{\max }(N)$ be the mode of maximum value, that is, the most
probable maximum value. It obeys an equation

\[
\left. \varphi _N^{^{\prime }}(x)\right| _{x=X_{\max }}=0 
\]

Since $p_{\alpha ,1}(x)\propto x^{-1-\alpha }$ for large $x$, we may give
the following estimate for the mode:

\begin{equation}
X_{\max }(N)\propto N^{1/\alpha }\;,\;0<\alpha <2.  \eqnum{19}
\end{equation}

With the help of Eq.(19) we are able to roughly estimate the diverging
statistical moment of the sequence $\{X(t)\}$, $t=1,2,...,N$, as

\begin{equation}
\int\limits_0^{X_{\max }(N)}dX\,X^qp_{\alpha ,1}(X)\propto X_{\max
}^{q-\alpha }\propto N^{q/\alpha -1},\;\,\,\,q>\alpha .  \eqnum{20}
\end{equation}

Now we turn to the structure function. It can be written as

\begin{equation}
S_q(\tau )=\left\langle \left| \Delta L_\alpha (\tau )\right|
^q\right\rangle =\int\limits_{-\infty }^\infty d\Delta L_\alpha \left|
\Delta L_\alpha \right| ^qp_{\alpha ,1}(\Delta L_\alpha ,\tau ),  \eqnum{21}
\end{equation}

where $\Delta L_{\alpha } (t)\equiv L_{\alpha } (t+\tau )-L_{\alpha } (t)$ ,
and the probability density is given by Eqs. (1), (2).

We may introduce a stochastic variable $\xi $, such that $\Delta L_\alpha
(\tau )=\xi \tau ^{1/\alpha }$ , and rewrite $S_q$,

\begin{equation}
S_q=\tau ^{q/\alpha }\int\limits_{-\infty }^\infty d\xi \;\xi ^qp_{\alpha
,1}(\xi ),  \eqnum{22}
\end{equation}

where the probability density for a new variable is given by Eqs. (13), (14).

To estimate the integral in Eq.(22), we use Eq.(20) with an important note
that here $N$ is equal to $T/\tau $, $T$ is the total length of the sample.
Therefore, the integral can be estimated as $(T/\tau )^{q/\alpha -1}$, and

\begin{equation}
S_q\propto \tau \;T^{q/\alpha -1},\quad q>\alpha .  \eqnum{23}
\end{equation}

Thus, the effects of a limited sample size manifest itself in Eq.(23), which
replaces ''theoretical infinity'' in Eq.(5) for $q>\alpha $ . A particular
case is the second order structure function, for which we have ''pseudo -
Gaussian'' relation (see Eq.(8)),

$S_2^{1/2}\propto \tau ^{1/2}$ [33]. The linear $\tau $- dependence of the $%
q $ - th order structure function was found in [34].

Equation (23) points to violation of self - affinity of the oLm. Indeed,
self - affinity implies that the - exponent of the structure function
depends linearly on $q$, see Eqs. (4), (5). On the contrary, the rough
estimate of the finite sample effects demonstrates that the exponent does
not depend on $q\,$at $q$ $>\alpha $ . Thus, in experimental data processing
one may expect that the exponent smoothly changes its slope, thus giving
rise to a convex curve. Such convexity seems as an indication of multi -
affinity, see Ref. [35]. However, in our case this behavior of the exponent
is stipulated not by ''intrinsic'' reason, but instead by the influence of a
finite size of a sample of the self - affine process. That is why we call
this effect ''spurious multi - affinity''.

Now we consider the finite size effects for the range of the oLm. In the
empirical rescaled range analysis, that is, at experimental data processing
or in numerical simulation the range of the random process is divided by the
standard deviation (that is, the square root of the second moment) for the
sequence of increments,

\[
\sigma _2=\left( \frac 1\tau \sum\limits_{t=1}^\tau \left( X(t)\right)
^2\right) ^{1/2} 
\]

after subtraction of a linear trend. This procedure, called the Hurst
method, or the method of normalized range, ''smoothes'' the fluctuations of
the range on different segments of time series, and is used in a great
variety of applications [27]. H. E. Hurst was the first [36] who has
collected large statistical material relating to water levels and other
phenomena, which indicates that the observed normalized range do not
increase (as it is expected for the ordinary Brownian motion) like the
square root of the observational period $\tau $, but instead like a higher
power. However, the Hurst method is not satisfactory for the oLm because of
the infinity of the theoretical value of standard deviation. What one should
expect when estimating the normalized range of the oLm from the finite
sample size ? Since $\left\langle X^2\right\rangle \propto \tau ^{2/\alpha
-1}$ , see Eq. (20), then

$\sigma _{2} \propto \tau ^{1/\alpha -1/2} $ , whereas $R(\tau )\propto \tau
^{1/\alpha } $ , and, thus,

\begin{equation}
\frac{R(\tau )}{\sigma _2}\propto \tau ^{1/2}.  \eqnum{24}
\end{equation}

Therefore, we conclude, that at the empirical rescaled range analysis the
Hurst method gives ''spurious'', ''pseudo - Gaussian'' time dependence.

In order to get the correct exponent $1/\alpha $, and, at the same time, to
smooth fluctuations of the range, we propose to modify the Hurst method by
exploiting the $\alpha $ - th root of the $\alpha $- th moment instead of
standard deviation, that is,

\begin{equation}
\sigma _\alpha =\left( \frac 1\tau \sum\limits_{t=1}^\tau \left| X(t)\right|
^\alpha \right) ^{1/\alpha }.  \eqnum{25}
\end{equation}

Since it has only weak logarithmic divergence with the number of terms in
the sum increasing, then one has

\begin{equation}
\overline{\left( \frac{R(\tau )}{\sigma _\alpha }\right) }\propto \tau ^H, 
\eqnum{26}
\end{equation}

where $H\approx 1/\alpha $ is the Hurst index for the oLm with the Levy
index , and the bar denotes averaging over the number of segments (having
the length ) of the sample path.

The expediency of using modified Hurst method when studying the Levy motion
can be explained as follows. In general case, the Hurst exponent in Eq.(4)
contains information not only on the Levy index of the increment
distribution, but also on long - time correlations between the increments.
In case of the ordinary Levy motion the correlations between non -
overlapping increments are absent, and the Hurst index is equal $1/\alpha $.
If correlations exist, the Hurst index for the Levy motion differ from $%
1/\alpha $, and this circumstance leads to violation of the ''$1/\alpha $
laws'' for the structure function and for the range. Therefore, when
treating experimental data, it seems expedient to estimate with the use of
increment distribution (there are different methods for estimating
parameters of stable distributions, see, e.g., [37]) and, then by testing ''$%
1/\alpha $ laws'' for the structure function and for the range, to get
information about the presence (or absence) of long - time correlations.

In the next Section we verify numerically the self - affine properties and
finite sample effects with the help of approximation described in Sec. 3.

\section{ Self - affine properties and finite sample effects in numerical
simulation}

We study numerically $\tau $- and $T$ - dependence of the $q$ - th order
structure function,

\begin{equation}
S_q\propto \tau ^{\mu (q)}T^{a(q)},  \eqnum{27}
\end{equation}

where, as before, is the time argument of the structure function, and $T$ is
the sample size. According to the analytical estimates

\begin{equation}
\mu (q)=\left\{ 
\begin{array}{c}
q/\alpha ,\quad q<\alpha \quad , \\ 
1,\quad q\geq \alpha \quad ,
\end{array}
\right.  \eqnum{28}
\end{equation}

whereas

\begin{equation}
a(q)=\frac q\alpha -1,  \eqnum{29}
\end{equation}

see Eqs. (5) and (23).

We first study $\mu \,$vs in the relation at $q$ less than $\alpha $. In
Fig.3 a typical example is depicted by crosses at fixed $q=1/2$. The $%
q/\alpha $ curve is shown by primes. One can be convinced himself that the $%
1/\alpha $ law is well confirmed at q smaller than the smallest Levy index
in numerical simulation. Also in the figure $\mu \,$vs $\alpha $ is depicted
by black points at $q=2$. It is shown that the second order structure
function, being estimated from a finite sample, leads to the spurious
''pseudo - Gaussian'' behavior. At the clarifying inset we show $S_q$ vs $%
\tau \,$in a log - log scale with $q$ and being fixed. The exponent $\mu \,$%
in the main figure is obtained for the fixed $\alpha \,$as a slope of a
straight line at the inset.

In Fig. 4a, b we plot $\mu \,$vs $q$ for the Levy index 1.2 and 1.7,
respectively. The analytical estimate, see Eq. (28), is shown by solid line
for $q<\alpha $ and by dotted line for $q>\alpha $ . The results of
simulation are shown by black points. The arrows indicate the value $%
q=\alpha $ , at which the bend of theoretical curves occur. It is shown that
the results of simulation are well fitted by the analytical curves.

Figure 5 demonstrates $S_q$ versus sample size $T$ in a log - log scale. The
Levy index $\alpha \,$is equal 1.2. The dotted lines indicate $S_q$ vs $T$
at $q=\alpha $ (horizontal line), $q=2.5\alpha $ and $4\alpha $ (sloping
lines). The black points indicate simulation results (4 points for each $q$%
). As in Fig. 4, we see a good agreement between experimental results and
analytical estimates. It implies, that Eq. (23), seeming very rough
estimate, nevertheless, accounts for the finite sample size effects for the
structure function of the oLm.

Figure 6 demonstrates the application of the modified Hurst method to the
sample path of the approximation to the oLm with the Levy index $\alpha =1$.
In Fig. 6a fluctuations of the range (thin curve) and those of $\sigma
_{\alpha \,}$(thick curve) are shown for the case, when the total length of
the sample is divided into 64 segments, each of $\tau =16$ lengthwise. Below
the variations of the ratio $R/\sigma _\alpha $ are depicted. It is shown
that fluctuations of the ratio is much smaller than those of the span. This
circumstance justify the use of the ratio in the empirical analysis. In Fig.
6b the rescaled span, see Eq. (26) is depicted versus the time interval by
black points in a log - log scale. The slope of the solid line is equal to $%
H=0.9$.

In Fig.7 the Hurst index obtained with using Eq.(26) is depicted by crosses,
whereas the curve $1/\alpha $ is indicated by primes. By comparing Fig.7
with Fig.3 one can see that the ''$1/\alpha $ law'' is better fulfilled for
the structure function of the simulated process than for the range. We also
note that the same conclusion follows for ''$\tau ^H$ laws'' for fractional
Brownian motion [15]. A more detailed discussion of this problem require the
use of the theory of extremes for the processes with stationary increments.
This is beyond the scope of our paper. However, we point readers attention
on the Hurst index which is obtained with the use of ''traditional'' ratio $%
R/\sigma _2$ and is shown by black points in Fig.7. In accordance with the
discussion of Sec.4, it can be clearly seen, that the standard deviation,
being used in empirical analysis of the oLm, ''suppresses'' the variations
of the range, thus giving rise to the spurious value $H\approx 0.5\,$for all
Levy indexes. On the contrary, only ''smoothes'' the variations of the
range, thus leading to the (nearly) correct value $H=1/\alpha $.

\section{ Conclusions.}

The ubiquity of the Levy motion rises the problems related to the
experimental data processing. In particular, the fat tails of the stable
distributions allows one to suggest that the effects of a finite sample size
play an important role. To study these effects, we propose an approximation
which is based on using Gnedenko limit theorem along with the method of
inversion. It allows us to simulate the ordinary Levy motion and study the
finite sample effects when estimating structure functions and the range. The
results of simulation are in a quantitative agreement with theoretical
estimates. In particular, we find, that the second order structure function
being estimated from the ordinary Levy motion sample paths as well as the
Hurst method both lead to spurious ''pseudo - Gaussian'' time dependencies.
We also propose to modify the Hurst method of rescaled range analysis in
order to avoid ''pseudo - Gaussian'' relation.

At the end we note that introduction of correlations into the sequence of
i.i.d. stably distributed random variables (with the help of the method used
in Ref. [15] for the sequence of i.i.d. Gaussian variables) allows us to
study the finite sample effects for fractional Levy motion.

Acknowledgements

This work was supported by the National Academy of Science of Ukraine, the
Project ``Chaos-2`` and by the INTAS Program, Grants LA - 96 - 02 and 98 -
01.

REFERENCES

1. P. Levy, Theorie de l'Addition des Variables Aleatories (Gauthier -
Villiers, Paris, 1937).

2. L. Bachelier, Annales Scientifiques de l'Ecole Normale Superieure III 17
(1900) 21.

3. N. Wiener, Journ. Math. Phys. Mass. Inst. Technology 2 (1923) 131 .

4. A. V. Skorokhod, Random Processes with Independent Increments ( in
Russian: Nauka, Moscow, 1964. Engl. transl.: Kluwer, Dordrecht, 1991);

5. D. A. Darling, Trans. Amer. Math. Soc. 83 (1956) 164.

6. R. M. Blummental, R. K. Getoor, Trans. Amer. Math. Soc. 95 (1960) 263;
Journ. Math. 4 (1960) 370.

7. G. Samorodnitsky, M. S. Taqqu, Stable non - Gaussian Random Processes
(Chapman \& Hall, New York, 1994).

8. B. V. Gnedenko, A. N. Kolmogorov, Limit distributions for Sums of
Independent Random Variables (in Russian: Izd-vo tekhniko - teor. lit - ry ,
Moskva, 1949; Engl. transl.:Addison Wesley, Reading, MA, 1954).

9. B. B. Mandelbrot, The Fractal Geometry of Nature (Freeman, New York,
1982).

10. . R. F. Voss, in: Fundamental Algorithms in Computer Graphics, edited by
R. A. Earnshaw (Springer-Verlag, Berlin, 1985), p.805.

11. Bar'yakhtar V., Gonchar V. et al., Ukrainian Journal of Physics , 38
(1993) 967.

12. B. B. Mandelbrot and J. R. Wallis, Water Resources Research 5 (1969) 228
.

13. S. Rambaldi, O. Pinazza, Physica A 208 (1994) 21.

14. O. Magre, M. Guglielmi, Chaos, Solitons and Fractals 8 (1997) 377.

15. . A. V. Chechkin, V. Yu. Gonchar, Preprint cond - mat/9902209 (accepted
to Chaos, Solitons \& Fractals).

16. . B. B. Mandelbrot, Journ. Business 36 (1963) 394;

17.B. Mandelbrot, Fractals and Scaling in Finance (Springer-Verlag, New
York, 1997).

18. . R. N. Mantegna, Physica A 179 (1991) 232.

19. . R. N. Mantegna, H. E. Stanley, Physica A 254 (1998) 77.

20. B. J. West and W. Deering, Phys. Reports 246 (1994) 1 .

21. M. F. Schlesinger, B. J. West, J. Klafter, Phys. Rev. Lett. 58 (1987)
1100.

22. M. F. Schlesinger, G. M. Zaslavsky, J. Klafter, Nature 363 (1993) 31; G.
M. Zaslavsky, Physica D 76 (1994) 110 ; M. F. Schlesinger (editor), G.
M.Zaslavsky, U. Frisch, Levy Flights and Related Topics in Physics:
Proceedings of the International Workshop Held at Nice, France, 27 - 30 June
1994 , Springer, Berlin, 1995).

23. J. - P. Bouchaud, A. Georges, Phys. Reports 195 (1990) 127 .

24. G. Zimbardo, P. Veltri, G. Basile, S. Principato, Phys. Plasmas 2 (1995)
2653.

25. D. Schertzer, S. Lovejoy, Multifractals and Turbulence: fundamentals and
applications (World Scientific, Singapore, 1993).

26. R. Gorenflo, G. De Fabritiis and F. Mainardi, Physica A269 (1999) 84.

27. J. Feder, Fractals (Plenum Press, New York, 1988).

28. B. B. Mandelbrot and van Ness, SIAM Review 10 (1968) 422.

29. W. Feller, Ann. Math. Stat. 22 (1951) 427.

30. R. N. Mantegna, Phys. Rev. E 5B (1994) 4677.

31. K. Umeno, Phys. Rev. E58 (1998) 2644.

32. M. G. Kendall, S. B. Babington, Random Sampling Numbers (Cambridge Univ.
Press, Cambridge, 1939).

33. A. V. Chechkin, V. Yu. Gonchar, Preprint cond - mat/9901064.

34. F. Schmitt, D. Scherzer, S. Lovejoy, Appl. Stochastic Models Data Anal.
15 (1999).

35. B. Mandelbrot, A. Fisher, L. Calvet, Cowles Foundation Discussion Paper
\# 1164 (1997).

36. H. E. Hurst, R. P. Black, and Y.M. Sinaika, Long Term Storage in
Reservoirs. An Experimental Study (Constable, London, 1965).

37. J. P. Nolan, Comm. In Stat. - Stochastic Models 13 (1997) 759; M.
Meerschaert and H.-P. Scheffler, J. Stat. Plann. Inference 71(1-2) (1998) 19.

\newpage
FIGURE CAPTIONS

Fig. 1. Probability densities (above) and their asymptotics (below) are
indicated for the sequences of random variables generated with the use of
the proposed numerical algorithm at the Levy indexes (a) $\alpha =1$, and
(b) $\alpha =1.5$. The probability densities and the asymptotics of the
stable laws are indicated by solid lines.

Fig. 2. Stationary sequences (thin lines) and ordinary Levy motion
trajectories (thick lines) at the different Levy indexes.

Fig. 3. Plots of the exponent in Eq.(12) versus the Levy index $\alpha \,$at 
$q=0.5$ (crosses) and $q=2$ (black points). The $q/\alpha $ curve is
depicted by dashed line. At the inset the structure function $S$ versus is
shown in a log - log scale at $\alpha =1$, $q=0.5$.

Fig. 4. The exponent $\mu \,$(see Eqs. (27), (28)) versus the order $q$ of
the structure function for (a) $\alpha =1.2$ and (b) $\alpha =1.7$. The
analytical estimate, see Eq. (28), is shown by solid line for $q<\alpha $
and by dotted line for $q>\alpha $ . The results of simulation are shown by
black points. The arrows indicate the value $q=\alpha $ , at which bend of
theoretical curve occurs.

Fig. 5. Structure function $S$ versus sample size $T\,$in a log - log scale
for the Levy index $\alpha =1$ and different values of $q$: $q=\alpha $
(horizontal line), $q=2.5\alpha $ and $q=4\alpha $ (sloping lines). The
black points indicate simulation results.

Fig. 6. (a) Variations of the range (thin curve), of $\sigma _{\alpha \,}\,$%
(thick curve) and of their ratio (below) at the different time intervals for
the oLm with $\alpha =1$. (b) Rescaled range, see Eq. (26), versus time
interval in log-log scale (black points). Solid line has a slope $H=0.9$.

Fig. 7. Plots of the Hurst exponent $H$ vs $\alpha $ estimated with the use
of Eq.(16) (crosses) and with the use of the traditional Hurst method (black
points). The $1/\alpha $ curve is depicted by dashed line.

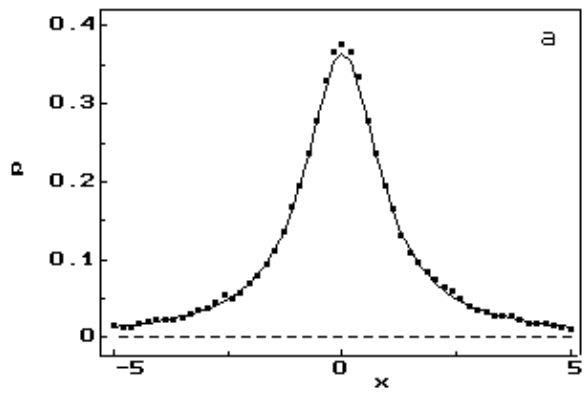
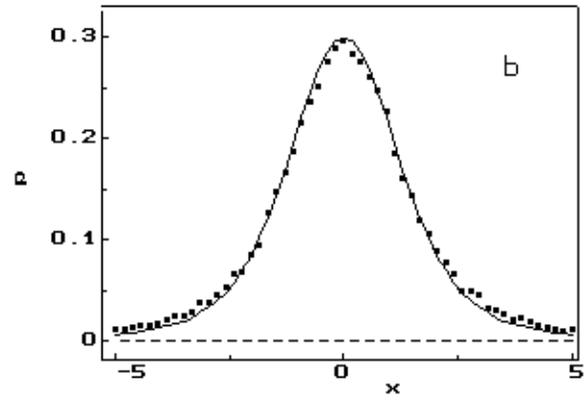
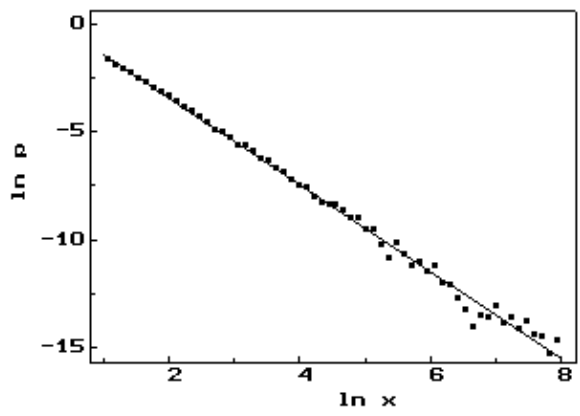
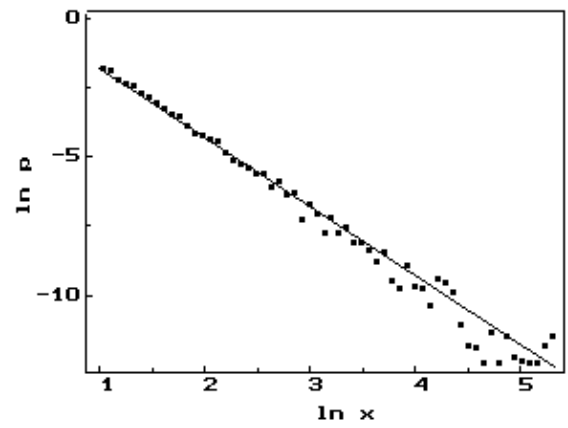
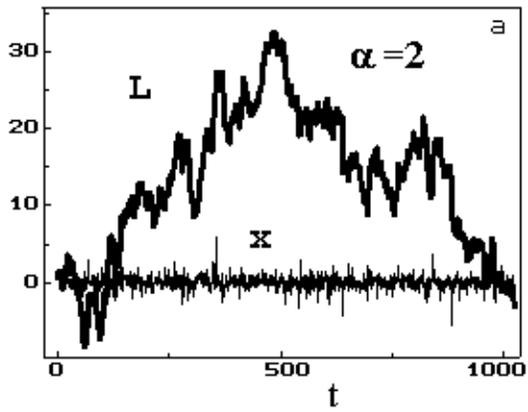
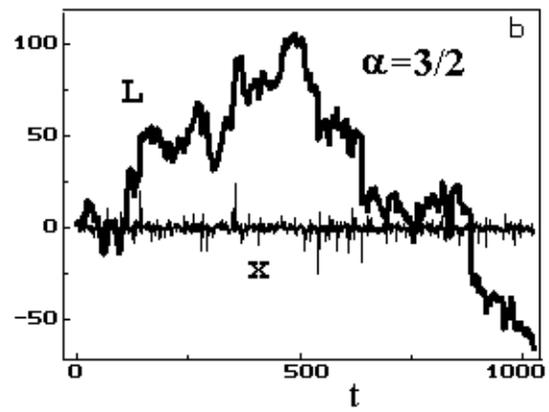
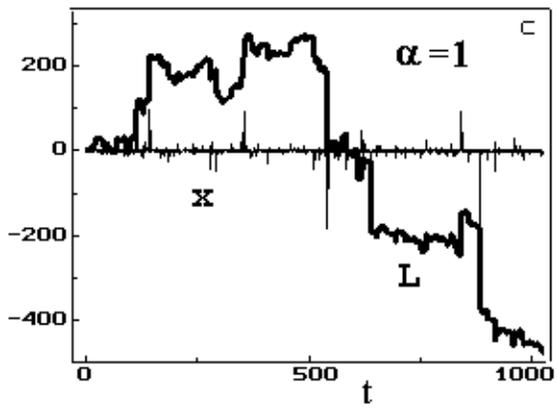
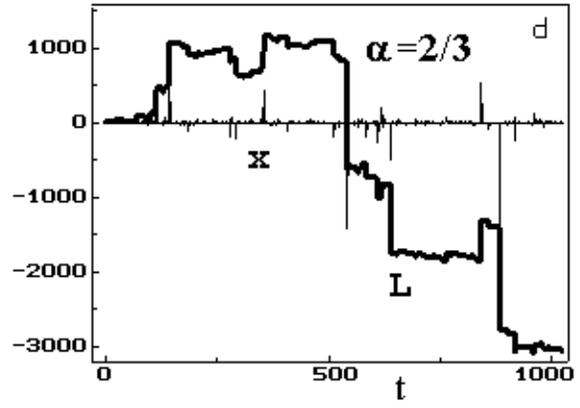
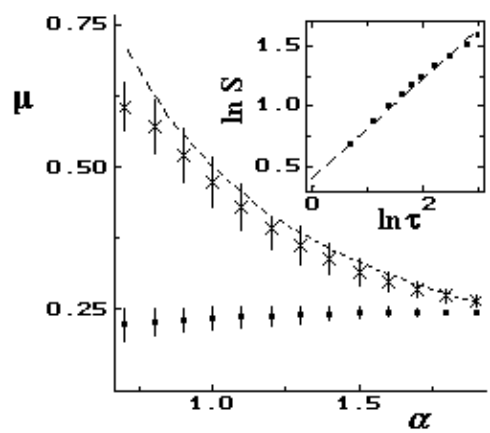
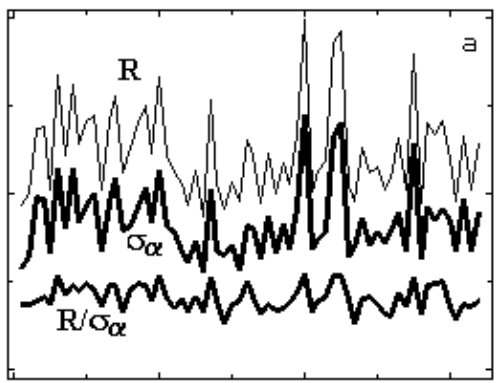
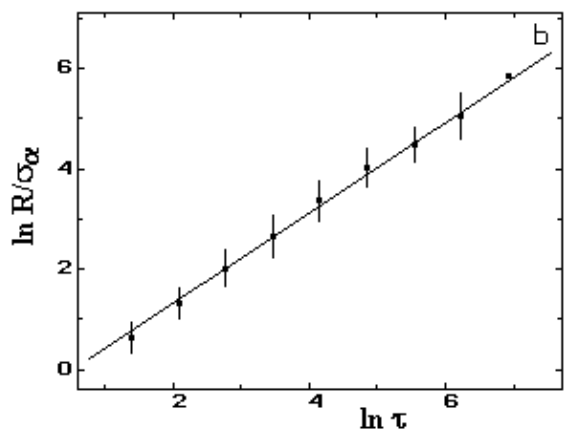
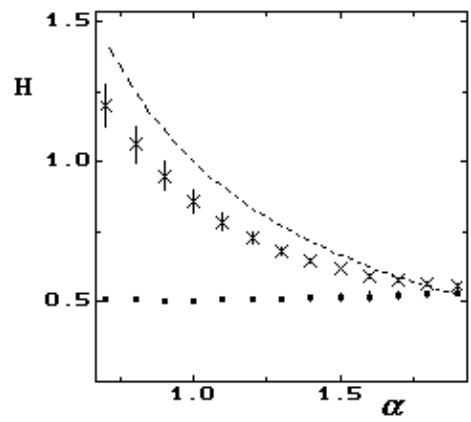
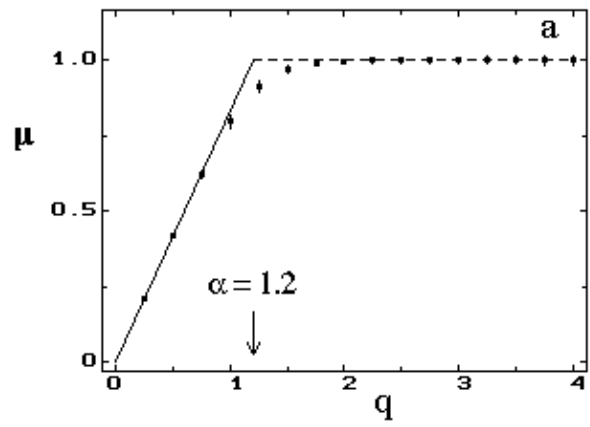
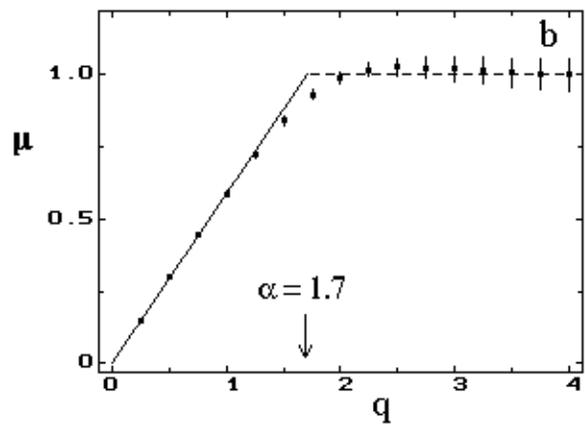
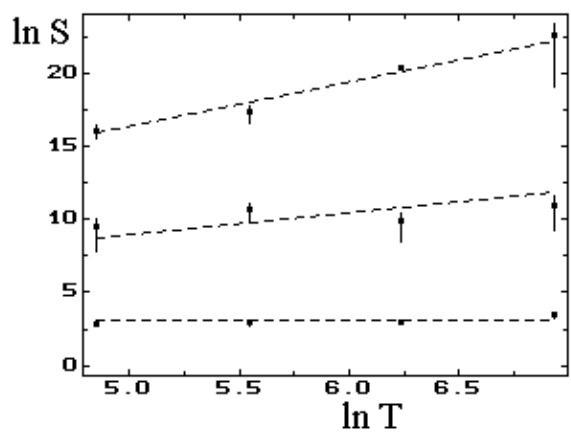
\end{document}